\UseRawInputEncoding
\documentclass[eqsecnum]{emulateapj}
\usepackage{natbib,amsmath}
\usepackage{epsfig}
\usepackage{hyperref}
\shortauthors{Gaur et al.}

\begin{document}

\shorttitle{Breaks in the X-ray spectra of high redshift blazars}

\title{Breaks in the X-ray spectra of high redshift blazars and the intervening medium}

\author{Haritma Gaur\altaffilmark{1}, Prashanth Mohan\altaffilmark{2}, Ashwani Pandey\altaffilmark{3} }
\altaffiltext{1}{Aryabhatta Research Institute of Observational Sciences (ARIES), Manora Peak, Nainital - 263 002, India; 
harry.gaur31@gmail.com}
\altaffiltext{2}{Shanghai Astronomical Observatory,
Chinese Academy of Sciences, 80 Nandan Road, Shanghai 200030, China; pmohan@shao.ac.cn}
\altaffiltext{3}{Indian Institute of Astrophysics, Block II, Kormangala, Bangalore--560034, India}

\begin{abstract}
\noindent
The flat spectrum radio quasars (FSRQs) are a sub-class of blazars characterised by prominent optical emission lines and a collimated large-scale jet along the observer line of sight. An X-ray spectral flattening has been reported in FSRQs (at relatively high redshifts), attributable to either absorption from gas along the line of sight or intrinsic jet based radiative processes. We study a sample of 16 high redshift FSRQs ($z$ of 1.1 -- 4.7; rest frame energy upto 50 keV) observed with {\it XMM-Newton} and {\it Swift} satellites spanning 29 epochs. The X-ray spectra are fit with a power law including free excess absorption and one multiplied by an exponential roll off to account for the intrinsic jet based processes. A statistical analysis is used to distinguish between these models to understand the origin of the spectral flattening. The model selection is unable to distinguish between them in ten of the sixteen FSRQs. Intrinsic jet based radiative processes are indicated in four FSRQs where we infer energy breaks consistent with their expectation from the external Compton scattering of low energy ambient photons. Two of the FSRQs indicate mixed results, supportive of either scenario, illustrating the difficulty in identifying X-ray absorption signatures. A clear detection can be employed to disentangle the relative contributions from the inter-galactic medium and the intra-cluster medium, the methodology of which is outlined and applied to the latter two sources.
\end{abstract}


\section{Introduction}
\label{sec:introduction}
\noindent
Blazars are the most energetic active galactic nuclei (AGN) with a strongly variable non-thermal broadband continuum emission
 \cite[e.g.][]{1995PASP..107..803U,1998MNRAS.299..433F} and are rare probes of activity in galactic nuclei and environmental properties, especially at high redshifts \cite[e.g.][]{2020NatCo..11..143A,2020ApJ...897..177P}. Their spectral energy distribution is double peaked with the lower ``synchrotron" peak originating from relativistic electrons in the magnetized jet, and the higher ``inverse Compton" peak originating from the Compton up-scattering of synchrotron photons by same population of relativistic electrons (synchrotron-self Compton, SSC) or external seed photons (external Compton, EC) sourced from the accretion disc, broad line region and torus components of the AGN \cite[e.g.][]{2010ApJ...716...30A,2018MNRAS.473.3638G}. 

Blazars can be classified into BL Lacs and flat spectrum radio quasars (FSRQs) based on the optical spectroscopic identification. The BL Lacs are characterized by a general lack of emission or absorption lines with equivalent widths $\leq$ 5 \AA, while the FSRQs have emission lines with significantly higher equivalent widths and a flat radio spectral index ($S_\nu \propto \nu^\alpha$, $\alpha \geq -0.5$) \cite[e.g.][]{1995PASP..107..803U} and are believed to be the counterparts of the strongly jetted (highly collimated to large scales) Fanaroff-Riley Type II radio galaxies \cite[e.g.][]{2002ApJ...564...86B} according to the radio-loud AGN unification scheme \cite[][]{1995PASP..107..803U}. The detection of a population of radio-loud $\gamma$-ray emitting narrow-line Seyfert 1 galaxies \cite[e.g.][]{2009ApJ...707L.142A} provides evidence for jet activity in terms of luminosities and radio component kinematics as well as in terms of their luminosity function which matches that of FSRQs, and host galaxy properties \cite[e.g.][]{2016A&A...591A..98B,2017FrASS...4....6F,2020Univ....6..136F}. The traditional classification then requires to be updated by a physically motivated scheme which may include jet power, accretion rate and black hole mass as defining parameters \cite[e.g.][]{2017FrASS...4....6F} such as an evolutionary blazar sequence \cite[e.g.][]{2002ApJ...564...86B, 2008MNRAS.387.1669G,2010MNRAS.405..387G}. 

The lower peak of the spectral energy distribution of FSRQs typically lies at $\leq 10^{15}$ Hz; their soft X-ray spectrum (0.1 $-$ 
10 keV energy range) can remain flat or indicate a curvature (convex or concave), with any transition in shape settling in the hard 
X-rays ($>$ 10 keV). For FSRQs at high redshifts, the observed soft X-ray emission approaches the relatively harder X-rays in the source 
rest frame as the corresponding frequencies are related by $\nu_{\rm rest} = (1+z)~\nu_{\rm obs}$. 

A relative flattening and spectral break is found in the X-ray spectra of some high redshift FSRQs \cite[e.g.][]{2006MNRAS.368..985Y}. This is attributed to either an absorption by a dense column of warm gas (with column density $N_{H} \geq 10^{22}$ cm$^{-2}$) in the vicinity of the host galaxy \cite[e.g.][]{1994ApJ...422...60E,2000ApJ...545..625Y,2001MNRAS.323..373F,2005MNRAS.358..432Y,2005MNRAS.364..195P} or to the intrinsic inverse-Compton emission profile in the hard X-rays \cite[e.g.][]{2001MNRAS.323..373F}. In the latter scenario, the flattening spectrum and break can be attributed to a cut-off in the lower energy tail of the electron energy distribution for SSC emission and to a sharply peaked external seed photon distribution for EC emission \cite[e.g.][]{2001MNRAS.323..373F,2004MNRAS.350L..67W,2007ApJ...665..980T}. 

The former scenario, i.e. a dense absorbing gas in the vicinity of the host galaxy is less plausible owing to the expectation that this can 
be quickly cleared by the powerful jet along the observer's line of sight in blazars \cite[e.g.][] {2016MNRAS.461..967M,2018A&A...616A.170A,
2019ApJ...882..130B}. A significant fraction of baryons in the observable Universe may be in the form of the photoionized and shock heated circum-galactic medium and inter-galactic-medium (IGM) \cite[e.g.][]{2006ApJ...650..560C,2012ApJ...759...23S}. The excess absorption and spectral flattening may then be attributable to an intervening warm IGM (at a temperature of $10^5 - 10^7$ K) along the line of sight \cite[e.g.][]{2018A&A...616A.170A}. This may not however present a complete picture as a fraction (possibly substantive) of the contribution can arise from the galaxy cluster environment \cite[e.g.][]{1991MNRAS.252...72W,1993ApJ...408...71W}, the intra-cluster medium (ICM). The ICM composition and thermodynamical state is shaped by complex mergers, AGN activity (outflows and radiation pressure) and winds from quiescent and star-burst galaxies that compose the galaxy cluster \cite[e.g.][]{1986RvMP...58....1S,2002ARA&A..40..539R,2002MNRAS.333..145N}. 

This interesting spectral regime can then probe the physical phenomena responsible for the spectral shape. X-ray spectroscopic studies in the direction of these blazars can help ascertain the relative contributions from the diffuse IGM \cite[e.g.][]{2011ApJ...734...26B,2019ApJ...882..130B} and discrete observational signatures (including contributions from the ICM) and aid in understanding if $N_H$ of the IGM evolves with redshift $z$ \cite[e.g.][]{2000MNRAS.316..234R,2018A&A...616A.170A}.

We study a sample of sixteen FSRQs at redshifts $z = 1.1 - 4.7$ observed with {\it XMM-Newton} and {\it Swift} (with rest frame energies from 0.4 keV -- 57 keV) spanning 29 epochs. The study is aimed at identifying spectral flattening in the X-ray spectra and investigating their origin in the excess absorption and intrinsic curvature scenarios and investigating the relative contributing components in the former. The analysis of data from both satellites serves as a means of confirmation or breaking degeneracies in the inference of the origin of the spectral flattening. The paper is structured as follows: in Section 2, we briefly describe the sample selection and data reduction; in Section 3, we present the results of the analysis which are then discussed.

\begin{table*}
\caption{Observation log of XMM-Newton X-ray data for FSRQs}
\begin{tabular}{lccccccc} \hline \hline
Blazar Name             & $\alpha_{2000.0}$& $\delta_{2000.0}$             & redshift  & Date of Obs. & Obs. ID  & $N^{**}_{H,G}$ &  Rest frame\\
                        &                  &                               &   $z$     & yyyy.mm.dd   &          & ($\times$ 10$^{20}$  cm$^{-2}$) & (keV)   \\ \hline
QSO B0014$$+$$810       & 00h17m08.0s      & $+$81$^{0}$35$^{'}$08$^{''}$  & 3.36  &  2001.08.23   & 0112620201  & 21.90   & (1.34-33.60) \\

RBS 315                 & 02h25m04.7s      & $+$18$^{0}$46$^{'}$48$^{''}$  & 2.69  &  2003.07.25   & 0150180101  & 16.80   & (1.08-26.90) \\
                        &                  &                               &       &  2013.01.13   & 0690900101  &       &\\
                        &                  &                               &       &  2013.01.15   & 0690900201  &       &\\

PMN 0525$-$3343         & 05h25m06.2s      & $-$33$^{0}$43$^{'}$05$^{''}$  & 4.40 &  2001.02.11   & 0050150101  & 2.54  & (1.76-44.01) \\
                        &                  &                               &       &  2003.02.14   & 0149500601  &       &\\
                        &                  &                               &       &  2003.08.08   & 0149501201  &       &\\
                        &                  &                               &       &  2003.02.24   & 0149500701  &       &\\
                        &                  &                               &       &  2003.03.06   & 0149500801  &       &\\
                        &                  &                               &       &  2003.03.16   & 0149500901  &       &\\
                        &                  &                               &       &  2001.09.15   & 0050150301  &       &\\

PKS 0524$-$460          & 05h25m18.1s      & $-$46:$^{0}$00$^{'}$21$^{''}$  & 1.48  &  2015.08.19   & 0762920501  &4.06   & (0.59-14.80) \\

PKS 0528$+$134          & 05h30m56.4s      & $+$13$^{0}$31$^{'}$55$^{''}$  & 2.06  &  2009.09.11   & 0600121601  & 38.40  & (0.82-20.60) \\
                        &                  &                               &       &  2009.09.14   & 0600121701  &       &\\

PKS 0537$-$286          & 05h39m54.3s      & $-$28$^{0}$39$^{'}$56$^{''}$  & 3.10 &  2000.03.19   & 0114090101  &2.39  & (1.24-31.04) \\
                        &                  &                               &       &  2005.03.20   & 0206350101  &       &\\

1ES 0836$+$710          & 08h41m24.4s      & $+$70$^{0}$53$^{'}$42$^{''}$  & 2.17 &  2001.04.12   & 0112620101  & 3.13  & (0.87-21.72) \\

RXJ 1028.6$-$0844        & 10h28m38.7s      & $-$08$^{0}$44$^{'}$38$^{''}$  & 4.28 &  2002.05.15   & 0093160701  & 5.31  & (1.71-42.76) \\
                        &                  &                               &       &  2003.06.13   & 0153290101  &       & \\

PKS 1127$-$145          & 11h30m07s        & $-$14$^{0}$49$^{'}$27$^{''}$  & 1.18 &  2002.07.01   & 0112850201  & 3.82  & (0.47-11.84) \\

PKS 1406$-$076          & 14h08m56.5s      & $-$07$^{0}$52$^{'}$27$^{''}$  & 1.49 &  2003.07.05   & 0151590101  & 2.77  & (0.60-14.94) \\
                        &                  &                               &       &  2003.08.10   & 0151590201  &       &\\

7C 1428$+$4218          & 14h30m23.7s      & $+$42$^{0}$04$^{'}$36$^{''}$  & 4.72  &  2002.12.09   & 0111260101  & 1.20   & (1.89-47.20) \\
                        &                  &                               &       &  2005.06.05   & 0212480701  &       & \\

GB 1508$+$5714          & 15h10m03.0s      & $+$57$^{0}$02$^{'}$44$^{''}$  & 4.30   &  2002.05.11   & 0111260201  & 1.63   & (1.72-43.00) \\

PBC J1656.2$-$3303    & 16h56m16.8s      & $-$33$^{0}$02$^{'}$12$^{''}$  & 2.40  &  2009.09.11   & 0601741401  & 33.50 & (0.96-24.00) \\

PKS 1830$-$211          & 18h33m39.9s      & $-$21$^{0}$03$^{'}$40$^{''}$  & 2.51 &  2004.03.24   & 0204580301  & 34.40  & (1.00-25.07) \\

PKS 2126$-$158          & 21h29m12.1s      & $-$15$^{0}$38$^{'}$42$^{''}$  & 3.27 &  2001.05.01   & 0103060101  & 6.01  & (1.31-32.68) \\

PKS 2149$-$306          & 21h51m55.3s      & $-$30$^{0}$27$^{'}$54$^{''}$  & 2.35 &  2001.05.01   & 0103060401  & 1.76  & (0.94-23.45) \\
\hline
\end{tabular}     \\
$^{*}$ Sources from our previous studies. \\
$^{**}$ $N_{H,G}$: Galactic absorption and values are taken from Willingale et al. (2013) \\
\label{obs_log}
\end{table*}

\begin{figure*}
\centering
\includegraphics[width=0.46\textwidth]{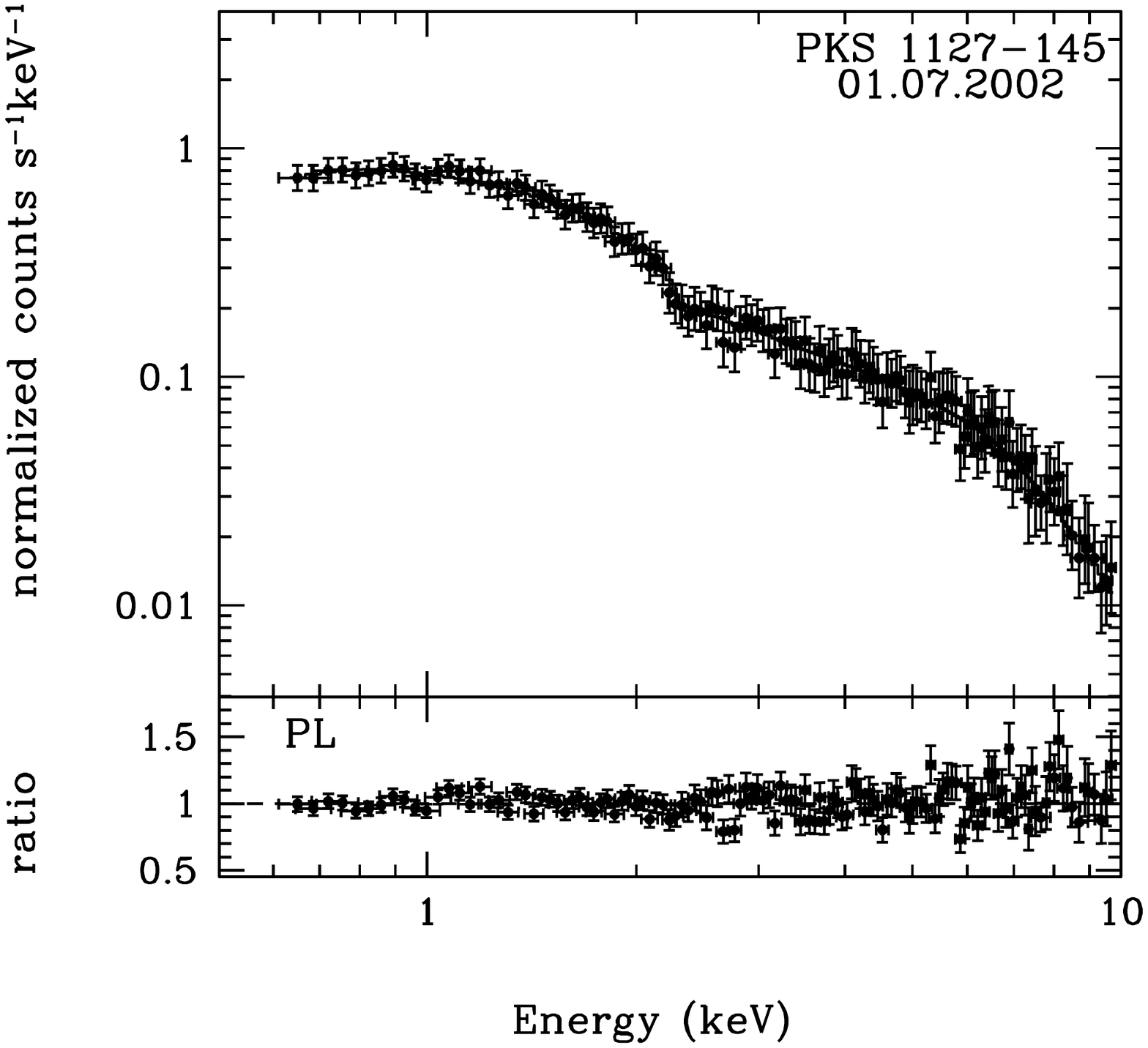}
\includegraphics[width=0.46\textwidth]{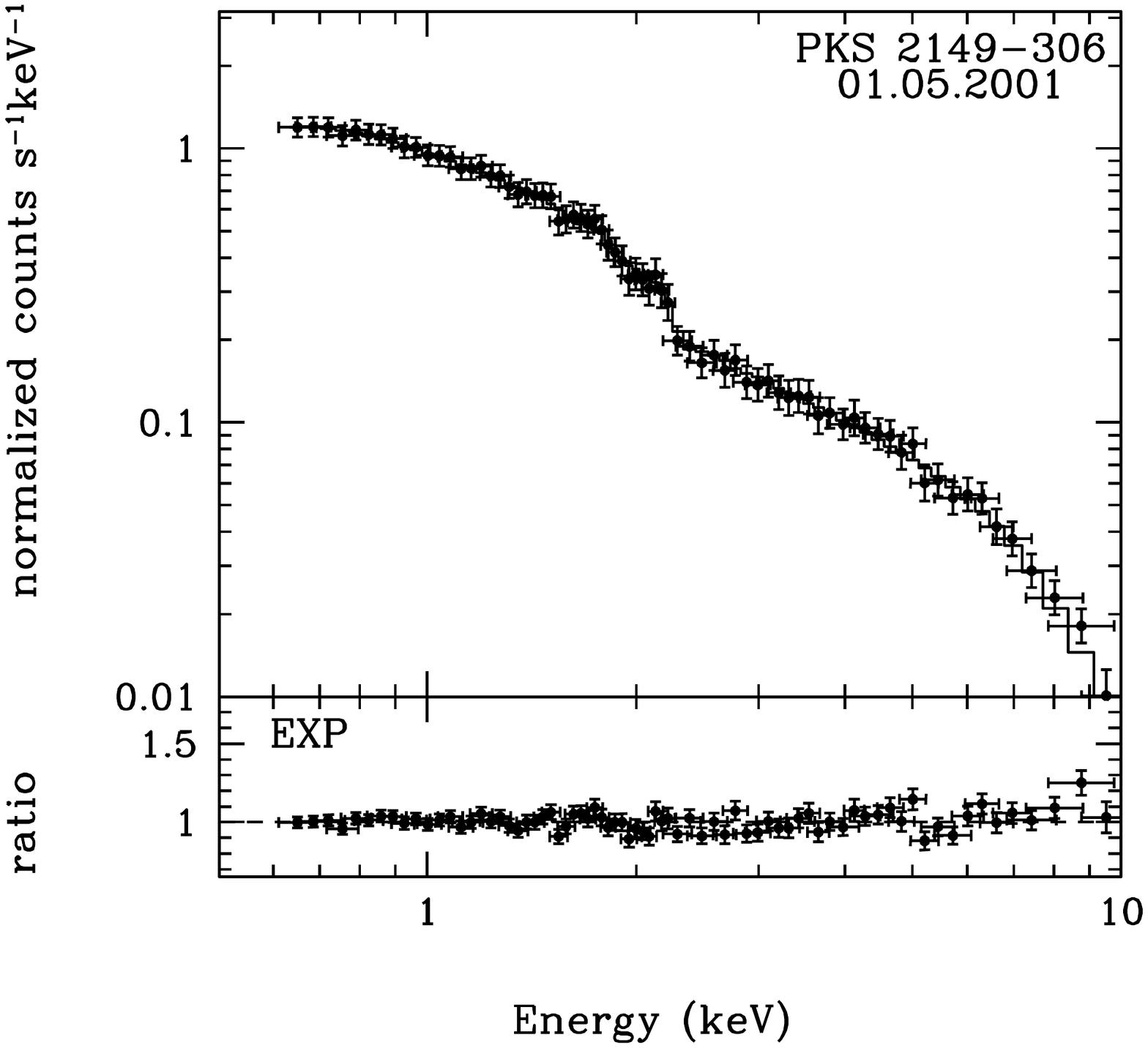}
\caption{Examples of the X-ray spectra fitted with the PL and EXP models and their corresponding ratios. The name and date of observation ({\it XMM-Newton}) of each blazar is also provided in upper right corner of each sub-figure.}
\label{fig1}
\end{figure*}

\section{Sample selection and data analysis}
\label{sec:data analysis}

\noindent
The study of \cite{2013ApJ...774...29E} present a large sample of 58 quasars with redshift $z \geq 0.45$ based on X-ray selection criteria, including a high photon count ($>$ 1800). We compiled a list of sources from this study with $z \geq 1$ to ensure that the ensuing X-ray spectra only consisted of the power law and relatively hard X-ray components; of these, we selected sources which also included a $\gamma$-ray detection\footnote{http://www.ssdc.asi.it/fermiagn/}, a criterion used to ensure the presence of a high energy spectral component. From this parent sample, we selected the FSRQs which were observed with the {\it XMM-Newton} and {\it Swift} satellites (in the energy range $0.3 - 10$ keV).

The sample in our study thus consists of 16 FSRQs spanning redshifts in the range of $1.1 - 4.7$. The sources, position (right ascension and declination), redshift, observation IDs with {\it XMM-Newton}, Galactic column density contribution and effective rest frame energy range are listed in Table \ref{obs_log}. 

The European Photon Imaging Camera (EPIC) pn data from {\it XMM-Newton} is used for the analysis as it is most sensitive and least affected by photon pile-up effects. We used the standard procedures described in \cite{2013Icar..226..186S} for the processing and analysis of the data. Only the single pixel events are extracted for the analysis. 
The X-ray spectra is extracted from a circular region centered on the source, with radius varying between 35--40 arcsec. The background to be subtracted is extracted from a circular region having the same radius, offset by about 180 arcsec from the source in the same chip set. Pile-up effects are checked for each observation using the science analysis system (SAS) task EPATPLOT. We found that the observations were not affected by pile up effects. The spectra were then rebinned to get a minimum of 30 counts in each energy bin. Due to the uncertainties in the calibration below 0.4 keV, we consider only the 0.4 $-$ 10 keV energy band for the present study. The photon redistribution matrix and the ancillary files are created with the SAS task {\it rmfgen} and {\it arfgen}, respectively.

The analysis of data from {\it Swift} was been limited to sources for which the product of the integrated flux and the exposure is greater than $1.0 \times 10^{−8}$ erg cm$^{-2}$ in the {\it Swift}-{\it X-ray telescope} ({\it XRT}). This allowed for the study of 13 FSRQs with high signal to-noise level, and consequently 
good-quality spectra. 
The data was analysed using HEASoft (version 6.28) software package and the updated calibration files (CALDB version 20200724). Using the standard {\it xrtpipeline} script, we obtained the cleaned and calibrated events files. We used data taken in Photon Counting (PC) mode in our analysis. A circular region of radius 20 pixels centered on the source was selected as source region while a circular region of radius 40 pixels free from source contamination was chosen as background region. We then extracted source and background light curves and spectra using the {\it xrtproducts} script with a bin size of 10 s in the energy range 0.3 - 10.0 keV. For the spectral studies, data are grouped using the grappha tool to have a minimum 20 counts per bin and the co-added spectra are fitted using XSPEC v. 12.10.1. Only those spectra are co-added where we did not find variability in the light curve.

\begin{table*}
\caption{Best fit spectral parameters for the power law (with excess absorption) and the powerlaw (with low energy exponential roll-off) for the high redshift FSRQs based on {\it XMM-Newton} observations.}
\noindent
\setlength{\tabcolsep}{0.011in}
\begin{tabular}{lcccccccr}\hline\hline

Source&Dateof&Model&$\Gamma$&$E_{f}$&log$_{10}$Flux&$N_{H}$&$\chi_{Red}^{2}$/dof&AIC\\
&observation&&&(keV)&&($10^{22}$cm$^{-2}$)&&\\\hline

QSO B0014$+$810    & 2001.08.23   &PL  &$  1.51_{-  0.02 }^{+  0.02}$ &  &$-11.31_{-  0.01 }^{+  0.01}$   &$<$  0.34     &1.07/139 & 154.73\\
   &   & EXP     &$  1.51_{-  0.02 }^{+  0.02}$  &$>$   0.14     &$-11.31_{-  0.01 }^{+  0.01}$      &  &1.07/ 139  &  154.73   \\
RBS 315            & 2003.07.25     &PL  &$  1.23_{-  0.01 }^{+  0.01}$  & &$-10.688_{-  0.004 }^{+  0.004}$   &$1.3_{-  0.2 }^{+  0.2}$       &1.28/163 & 214.64\\

           &   & EXP     &$  1.26_{-  0.02 }^{+  0.02}$  &$  0.36_{-  0.03 }^{+  0.02}$     &$-10.689_{-  0.004 }^{+  0.004}$      &  &1.22/ 163  & {\em 204.86}   \\
       &2013.01.13   &PL  &$  1.44_{-  0.01 }^{+  0.01}$  & &$-10.723_{-  0.003 }^{+  0.003}$   &$1.6_{-  0.1 }^{+  0.1}$       &1.31/169 & 227.39\\
                  &    & EXP     &$  1.48_{-  0.01 }^{+  0.01}$  &$  0.40_{-  0.01 }^{+  0.01}$     &$-10.724_{-  0.003 }^{+  0.003}$      &  &1.13/ 169  & {\em 196.97}   \\

       & 2013.01.15  &PL  &$  1.42_{-  0.01 }^{+  0.01}$  & &$-10.726_{-  0.002 }^{+  0.002}$   &$1.7_{-  0.1 }^{+  0.1}$       &1.53/170 & 266.10\\
                   &     & EXP     &$  1.45_{-  0.01 }^{+  0.01}$  &$  0.40_{-  0.01 }^{+  0.01}$     &$-10.728_{-  0.002 }^{+  0.002}$      &  &1.29/ 170  & {\em 225.30}    \\

      PMN 0525$-$3343    & 2001.02.11  &PL  &$  1.6_{-  0.1 }^{+  0.2}$       &  &$-11.96_{-  0.05 }^{+  0.04}$  &$ < 3.3$      &1.08/30 & 38.40\\
    &    & EXP     &$  1.7_{-  0.2 }^{+  0.2}$  &$< 0.4$     &$-11.96_{-  0.05 }^{+  0.05}$      &  &1.07/  30  &  38.10     \\
         &2001.09.15  &PL  &$  1.61_{-  0.05 }^{+  0.05}$     &  &$-11.87_{-  0.02 }^{+  0.02}$  &$  1.4_{-  0.9 }^{+  1.0}$      &0.89/83 & 79.87\\
                   &   & EXP     &$  1.63_{-  0.06 }^{+  0.06}$  &$  0.23_{-  0.10 }^{+  0.07}$     &$-11.87_{-  0.02 }^{+  0.02}$      &  &0.89/  83  &  79.87    \\
       & 2003.02.14   &PL  &$  1.62_{-  0.07 }^{+  0.07}$    &  &$-11.85_{-  0.02 }^{+  0.02}$  &$  2.1_{-  1.3 }^{+  1.4}$      &0.98/62 & 66.76\\
                   &  & EXP     &$  1.66_{-  0.08 }^{+  0.08}$  &$  0.3_{-  0.1 }^{+  0.1}$     &$-11.85_{-  0.02 }^{+  0.02}$      &  &0.96/  62  &  65.52    \\
       &2003.02.24   &PL  &$  1.62_{-  0.07 }^{+  0.07}$    &  &$-11.84_{-  0.02 }^{+  0.02}$  &$  1.5_{-  1.3 }^{+  1.3}$      &0.92/62 & 63.04\\
                   &  & EXP     &$  1.64_{-  0.08 }^{+  0.09}$  &$  0.24_{-  0.15 }^{+  0.09}$     &$-11.85_{-  0.02 }^{+  0.02}$      &  &0.92/  62  &  63.04    \\
       &2003.08.08  &PL  &$  1.60_{-  0.08 }^{+  0.08}$    &  &$-11.78_{-  0.03 }^{+  0.03}$  &$< 2.7$      &1.03/54 & 61.62\\
                  &   & EXP     &$  1.62_{-  0.09}^{+  0.10}$  &$ <0.3$     &$-11.78_{-  0.03 }^{+  0.03}$      &  &1.03/  54  &  61.62    \\
                & 2003.03.06  &PL  &$  1.57_{-  0.08 }^{+  0.08}$   &  &$-11.81_{-  0.03 }^{+  0.03}$  &$ <  2.3$      &1.03/53 & 60.59\\
                  &  & EXP     &$  1.59_{-  0.09 }^{+  0.10}$  &$ < 0.3$     &$-11.81_{-  0.03 }^{+  0.03}$      &  &1.03/  53  &  60.59    \\
                  &2003.03.16  &PL  &$  1.62_{-  0.09 }^{+  0.10}$   &  &$-11.83_{-  0.03 }^{+  0.03}$  &$  2.1_{-  1.8 }^{+  1.9}$      &1.27/44 & 61.88\\
                   &     & EXP     &$  1.7_{-  0.1 }^{+  0.1}$  &$  0.29_{-  0.16 }^{+  0.11}$     &$-11.83_{-  0.03 }^{+  0.03}$      &  &1.26/  44  &  61.44    \\
PKS 0524$-$460     & 2015.08.19   &PL  &$  1.83_{-  0.05 }^{+  0.05}$ &  &$-11.99_{-  0.02 }^{+  0.02}$   &$>0.82$       &7.32/96 & 708.72\\
    &     & EXP     &$  1.83_{-  0.05 }^{+  0.05}$  &$<0.82$     &$-11.99_{-  0.02 }^{+  0.02}$      &  &7.32/  96  &  708.72   \\
PKS 0528$+$134     & 2009.09.11   &PL  &$  1.54_{-  0.06 }^{+  0.07}$   &  &$-11.80_{-  0.02 }^{+  0.02}$  &$  1.9_{-  0.6 }^{+  0.7}$      &1.14/95 & 114.30\\
                &     & EXP     &$  1.59_{-  0.08 }^{+  0.08}$  &$  0.53_{-  0.09 }^{+  0.09}$     &$-11.81_{-  0.02 }^{+  0.02}$      &  &1.13/  95  &  113.35   \\
    & 2009.09.14     &PL  &$  1.6_{-  0.1 }^{+  0.1}$  &  &$-11.78_{-  0.03 }^{+  0.03}$  &$  3.3_{-  1.3 }^{+  1.5}$      &0.82/53 & 49.46\\
                   &    & EXP     &$  1.6_{-  0.1 }^{+  0.1}$  &$  0.6_{-  0.2 }^{+  0.1}$     &$-11.78_{-  0.03 }^{+  0.03}$      &  &0.85/  53  &  51.05    \\
PKS 0537$-$286     & 2000.03.19   &PL  &$  1.29_{-  0.02 }^{+  0.03}$   &  &$-11.45_{-  0.01 }^{+  0.01}$  &$<0.21$      &1.05/132 & 144.60\\
 &     & EXP     &$  1.30_{-  0.02 }^{+  0.02}$  &$>  0.02$     &$-11.43_{-  0.01 }^{+  0.01}$      &  &1.05/ 132  &  144.60   \\
        &2005.03.20   &PL  &$  1.24_{-  0.03 }^{+  0.03}$    &  &$-11.22_{-  0.01 }^{+  0.01}$  &$<  0.7$      &1.22/119 & 151.18\\
                   &    & EXP     &$  1.24_{-  0.04 }^{+  0.04}$  &$< 0.2$     &$-11.22_{-  0.01 }^{+  0.01}$      &  &1.22/ 119  &  151.18   \\
      1ES 0836$+$710     & 2001.04.12  &PL  &$  1.31_{-  0.01 }^{+  0.01}$    &  &$-10.311_{-  0.003 }^{+  0.003}$  &$ >  0.06$      &1.20/167 & 206.40\\
     &   & EXP     &$  1.31_{-  0.01 }^{+  0.01}$  &$<  0.1$     &$-10.311_{-  0.003 }^{+  0.003}$      &  &1.20/ 167  &  206.40    \\
RXJ 1028.6-0844    & 2002.05.15   &PL  &$  1.18_{-  0.08 }^{+  0.09}$ &  &$-11.71_{-  0.03 }^{+  0.03}$   &$< 3.0$       &0.95/50 & 53.50\\
    &    & EXP     &$  1.18_{-  0.08 }^{+  0.10}$  &$ < 0.3$     &$-11.71_{-  0.03 }^{+  0.03}$      &  &0.96/  50  &  54.00       \\
         &2003.06.13  &PL  &$  1.46_{-  0.05 }^{+  0.05}$  & &$-11.89_{-  0.02 }^{+  0.02}$   &$2.7_{-  1.1 }^{+  1.1}$       &0.78/103 & 86.34\\
                   &     & EXP     &$  1.50_{-  0.06 }^{+  0.06}$  &$  0.33_{-  0.07 }^{+  0.06}$     &$-11.89_{-  0.02 }^{+  0.02}$      &  &0.77/ 103  &  85.31    \\

PKS 1127$-$145     & 2002.07.01        &PL  &$  1.30_{-  0.03 }^{+  0.03}$      &  &$-11.08_{-  0.01 }^{+  0.01}$  &$  0.14_{-  0.02 }^{+  0.02}$      &1.03/143 & 153.29\\
     &    & EXP     &$  1.32_{-  0.03 }^{+  0.03}$  &$  0.40_{-  0.03 }^{+  0.03}$     &$-11.08_{-  0.01 }^{+  0.01}$      &  &1.05/ 143  &  156.15   \\
PKS 1406$-$076     & 2003.07.05  &PL  &$  1.58_{-  0.06 }^{+  0.08}$     &  &$-12.20_{-  0.03 }^{+  0.03}$  &$< 0.06$ &1.10/44 & 54.40\\
     &     & EXP     &$  1.58_{-  0.07 }^{+  0.08}$  &$<0.82$     &$-12.20_{-  0.03 }^{+  0.03}$      &  &1.10/  44  &  54.40     \\
       &2003.08.10  &PL  &$  1.5_{-  0.1 }^{+  0.2}$    &  &$-12.18_{-  0.05}^{+  0.04}$  &$< 0.40$      &1.03/24 & 30.72\\
                   &     & EXP     &$  1.5_{-  0.1 }^{+  0.2}$  &$  > 0.34$     &$-12.18_{-  0.05 }^{+  0.03}$      &  &1.03/  24  &  30.72    \\

7C 1428$+$4218     & 2002.12.09  &PL  &$  1.92_{-  0.07 }^{+  0.07}$ &  &$-11.61_{-  0.02}^{+  0.02}$   &$< 0.66$ &1.00/45 & 51.00\\
     &     & EXP     &$  1.92_{-  0.05 }^{+  0.08}$  &$>  0.07$     &$-11.64_{-  0.02 }^{+  0.02}$      &  &1.00/  45  &  51.00       \\
       & 2005.06.05  &PL  &$  1.63_{-  0.03 }^{+  0.03}$ &  &$-11.71_{-  0.01 }^{+  0.01}$   &$<0.29$       &1.03/97 & 105.91\\
                   &     & EXP     &$  1.63_{-  0.03 }^{+  0.03}$  &$> 0.09$     &$-11.73_{-  0.01 }^{+  0.01}$      &  &1.03/  97  &  105.91   \\
GB 1508$+$5714     & 2002.05.11  &PL  &$  1.6_{-  0.1 }^{+  0.1}$ &  &$-12.29_{-  0.04 }^{+  0.04}$   &$<  3.2$       &1.32/39 & 57.48\\
     &    & EXP     &$  1.57_{-  0.08 }^{+  0.16}$  &$ <  0.3$     &$-12.29_{-  0.04 }^{+  0.03}$      &  &1.33/  39  &  57.87    \\
PBC J1656.2$-$3303 & 2009.09.11   &PL  &$  1.17_{-  0.03 }^{+  0.03}$ &  &$-11.20_{-  0.01 }^{+  0.01}$   &$7.5_{-  0.6 }^{+  0.6}$       &1.27/133 & 174.91\\
 &     & EXP     &$  1.29_{-  0.04 }^{+  0.04}$  &$  0.89_{-  0.03 }^{+  0.03}$     &$-11.21_{-  0.01 }^{+  0.01}$      &  &1.11/ 133  & {\em 153.63}   \\
PKS 1830$-$211     & 2004.03.24         &PL  &$  1.09_{-  0.02 }^{+  0.02}$     &  &$-10.828_{-  0.004 }^{+  0.004}$  &$  3.1_{-  0.2 }^{+  0.2}$      &2.00/165 & 336.00\\
     &     & EXP     &$  1.22_{-  0.03 }^{+  0.03}$  &$  1.13_{-  0.03 }^{+  0.03}$     &$-10.831_{-  0.004 }^{+  0.004}$      &  &1.54/ 165  & {\em 260.10 }   \\

PKS 2126$-$158     & 2001.05.01  &PL  &$  1.40_{-  0.02 }^{+  0.02}$     &  &$-10.92_{-  0.01 }^{+  0.01}$  &$  1.1_{-  0.2 }^{+  0.3}$      &1.01/150 & {\em 157.50}\\
     &    & EXP     &$  1.48_{-  0.04 }^{+  0.04}$  &$  0.24_{-  0.05 }^{+  0.05}$     &$-10.92_{-  0.01 }^{+  0.01}$      &  &1.07/ 150  &  166.50    \\

PKS 2149$-$306     & 2001.05.01         &PL  &$  1.44_{-  0.01 }^{+  0.01}$    &  &$-11.10_{-  0.01 }^{+  0.01}$  &$< 0.07$      &1.02/149 & 157.98\\
    &    & EXP     &$  1.44_{-  0.01 }^{+  0.01}$  &$>0.02$     &$-11.10_{-  0.01 }^{+  0.01}$      &  &1.02/ 149  &  157.98   \\

\hline

\hline
\end{tabular}\\
$\Gamma$: X-ray spectral index; $E_f$ (keV): is the folding energy of the exponential roll-off; 
logarithm of the flux (in ergs/sec/cm$^{2}$); $N_{H}$ (cm$^{-2}$): excess absorption; $\chi_{Red}^{2}$: reduced $\chi^2$ and degrees of freedom (dof); AIC: Akaike information criteria; italicised values denote the best fit model. 
\label{paratabxmm_2}
\end{table*}

\begin{table*}
\caption{Best fit spectral parameters for the power law (with excess absorption) and the power law (with low energy exponential roll-off) for the high redshift FSRQs based on {\it Swift - XRT} observations.}
\begin{tabular}{lccccccc} \hline \hline

Source&Model&$\Gamma$&$E_{f}$&log$_{10}$Flux&$N_{H}$&$\chi_{Red}^{2}$/dof&AIC\\
&&&(keV)&&($10^{22}$cm$^{-2}$)&&\\\hline \\

QSO B0014$$+$$81  &PL      &$  1.84_{-  0.09 }^{+  0.09}$    &      &$ -9.30_{-  0.02}^{+  0.02}$   &$  7.3_{-  2.1 }^{+  2.5}$     &0.94/  94 &   94.36       \\
  &EXP  &$2.0_{-0.1 }^{+0.1}$    & $0.7_{-0.1 }^{+0.1}$  &$ -9.30_{-0.02 }^{+0.02}$ &&0.85/  94   &{\em 85.90}  \\

RBS 315      &PL      &$  1.54_{-  0.07 }^{+  0.07}$    &      &$ -8.98_{-  0.02 }^{+  0.02}$   &$  3.9_{-  0.9 }^{+  1.0}$     &1.30/ 137 &    184.10      \\
    &EXP  &$1.62_{-0.08 }^{+0.09}$    & $0.60_{-0.07 }^{+0.07}$  &$ -8.98_{-0.02 }^{+0.02}$ &&1.30/ 137   &     184.10 \\
PKS 0524$-$460       &PL      &$  1.8_{-  0.2 }^{+  0.2}$    &      &$ -9.88_{-  0.05 }^{+  0.05}$   &$  1.4_{-  0.5 }^{+  0.6}$     &1.02/  23 &    29.46      \\
     &EXP  &$1.9_{-0.2 }^{+0.2}$    & $0.6_{-0.1 }^{+0.1}$  &$ -9.89_{-0.05 }^{+0.05}$ &&1.03/  23   &     29.69 \\

PKS 0528$+$134    &PL    &$  1.39_{-  0.09 }^{+  0.10}$    &   &$-11.38_{-  0.02 }^{+  0.02}$ & $  2.4_{-1.4 }^{+1.6}$    &0.89/  97        &   92.33       \\

  &EXP  &$1.5_{-0.1 }^{+0.1}$   &$0.6_{-0.2 }^{+0.2}$  &$-11.38_{-0.03 }^{+0.02}$    & &0.85/  97 &       {\em88.45}   \\

PKS 0537$-$286      &PL  &$  1.18_{-  0.05 }^{+  0.07}$     &   &$-11.45_{-  0.02 }^{+  0.02}$  & $ <0.82$    &0.98/ 121 &   124.58      \\
   &EXP  &$1.19_{-0.05 }^{+0.08}$   &$<0.25$  &$-11.45_{-0.02 }^{+0.02}$    & &0.98/ 121 &     124.58  \\
1ES 0836$+$710   &PL      &$  1.57_{-  0.03 }^{+  0.03}$    &   &$ -8.63_{-  0.01 }^{+  0.01}$ & $  1.8_{-0.2 }^{+0.2}$   &1.45/ 449 &    657.05     \\
 &EXP  &$1.65_{-0.03 }^{+0.03}$   &$0.53_{-0.02 }^{+0.02}$  &$ -8.64_{-0.01 }^{+0.01}$    & &1.43/ 449 &     658.30  \\
PKS 1127$-$145    &PL   &$  1.73_{-  0.06 }^{+  0.07}$   &   &$ -9.16_{-  0.02 }^{+  0.02}$ & $  0.23_{-0.03 }^{+0.04}$   &1.03/ 175 &   186.25      \\
   &EXP  &$1.77_{-0.07 }^{+0.07}$   &$0.61_{-0.05 }^{+0.05}$  &$ -9.16_{-0.02 }^{+0.02}$    & &1.04/ 175 &     188.00     \\
PKS 1406$-$076    &PL    &$  1.9_{-  0.2 }^{+  0.2}$ &   &$-10.11_{-  0.04}^{+  0.04}$ & $  0.9_{-0.5 }^{+0.6}$    &0.99/  35 &    40.65      \\
      &EXP  &$2.0_{-0.2 }^{+0.2}$   &$0.5_{-0.2 }^{+0.1}$  &$-10.11_{-0.04 }^{+0.04}$    & &0.99/  35 &     40.65   \\

7C 1428$+$4218     &PL      &$  1.7_{-  0.2 }^{+  0.3}$      &    &$ -9.71_{-  0.07 }^{+  0.07}$   &$< 12.8$     &1.54/  18 &   33.72       \\
      &EXP  &$1.8_{-0.3 }^{+0.3}$    & $<0.65 $  &$ -9.73_{-0.07 }^{+0.07}$ &&1.51/  18   &     33.18 \\

PBC J1656.2$-$3303  &PL      &$  1.6_{-  0.2 }^{+  0.2}$      &    &$ -9.28_{-  0.04 }^{+  0.03}$   &$ 15_{-  3 }^{+  4}$     &1.13/  52 &   64.76       \\

 &EXP  &$1.8_{-0.2 }^{+0.2}$    & $1.2_{-0.1 }^{+0.1}$  &$ -9.29_{-0.04 }^{+0.04}$ &&1.23/  52   &     69.96 \\
PKS 1830$-$211    &PL    &$  1.44_{-  0.07 }^{+  0.07}$      & &$ -8.89_{-  0.01 }^{+  0.01}$  & $  3.6_{-0.4 }^{+0.5}$   &1.23/ 232 &   291.36      \\
  &EXP  &$1.60_{-0.08 }^{+0.08}$   &$1.22_{-0.07 }^{+0.07}$  &$ -8.89_{-0.01 }^{+0.01}$    & &1.22/ 232 &     289.04  \\

PKS 2126$-$158     &PL    &$  1.74_{-  0.04 }^{+  0.04}$   & &$ -9.01_{-  0.01 }^{+  0.01}$  & $  6.3_{-0.7 }^{+0.7}$   &1.09/ 268 &    298.12     \\

   &EXP  &$2.30_{-0.09 }^{+0.09}$   &$1.4_{-0.1 }^{+0.1}$  &$ -9.03_{-0.01 }^{+0.01}$    & &1.06/ 268 & {\em 290.08}  \\
PKS 2149$-$306    &PL   &$  1.52_{-  0.03 }^{+  0.03}$       & &$ -8.75_{-  0.01 }^{+  0.01}$ & $  2.0_{-0.2 }^{+0.3}$    &1.25/ 334 &    423.50      \\
    &EXP  &$1.59_{-0.04 }^{+0.04}$   &$0.51_{-0.03 }^{+0.03}$  &$ -8.75_{-0.01 }^{+0.01}$    & &1.23/ 334 &{\em 416.82 } \\\hline

\end{tabular}
\label{paratabswift}
\end{table*}

\noindent
\subsection{Spectral fitting and analysis}

A power law (PL) with an excess absorption component is used to constrain the line of sight integrated column density $N_H$ (in excess of the Galactic contribution); the PL model offers a first estimate of parametric properties. 
The low energy portion of the X-ray spectrum i.e. $\le$ 2 keV shows a curvature due to a deficit of photons which is difficult to fit with a broken power-law model \citep{2007MNRAS.382L..82G}. Hence, the power-law model is multiplied by a roll-off term with a quadratic exponential form as employed in \cite{2009AdSpR..43.1036F}. The latter model is presumed to describe cases where the curvature may be attributable to the intrinsic electron distribution or the consequent synchrotron and inverse-Compton scattered photon energy distribution. The models and the associated parameters are: 

\begin{enumerate}

\item Power law with excess absorption (PL):
\begin{equation}
F(E) = k e^{-\left(N_{H,G}+N_{H}\right) \sigma (E)} E^{\Gamma},
\end{equation}
with free parameters including the normalization $k$, excess column density $N_H$ (in cm$^{-2}$) representing the contributions from the blazar host galaxy and the intervening IGM, and the spectral index $\Gamma$; $\sigma$ (in cm$^2$) is the energy dependent absorption cross section. 

\item Power law with exponential roll-off (EXP):
\begin{equation}
F(E) = k e^{-N_{H,G} \sigma (E)} E^{\Gamma} e^{-E^{2}_{f}/E^{2}},
\end{equation}
with free parameters including the normalization $k$, a spectral index $\Gamma$, and the folding energy of the exponential roll-off (which defines the low energy photon deficit) $E_{f}$ (in keV).

%
\end{enumerate}

All the spectra are fit with the {\it ztbabs} model in XSPEC which allows the addition of absorbing components at the redshift of the source \citep{2000ApJ...542..914W}. The Galactic absorption column density $N_{H,G}$ (cm$^{-2}$) is obtained from the survey by \cite{2013MNRAS.431..394W}. This includes the atomic gas column density $N_{HI}$ \citep{1997agnh.book.....H,2000A&AS..142...35A,2005A&A...440..767B,2005A&A...440..775K} and the molecular hydrogen column density $N_{H_{2}}$ \citep{1998ApJ...500..525S,2001ApJ...547..792D}. Abundances from \cite{2000ApJ...542..914W} and cross sections from \cite{1996ApJ...465..487V} are used. 
For blazars that are gravitationally lensed by an intervening galaxy hosting molecular gas \cite[PKS 1830-211; e.g.][]{1996Natur.379..139W} and by damped Lyman-$\alpha$ systems \cite[PKS 1127-145, 1ES 0836+710; e.g.][]{2005ARA&A..43..861W,2006A&A...453..829F} along the line of sight, additional absorbing components are placed at the appropriate redshifts.  

The X-ray spectra are fit using the $\chi^2$ method to estimate the free parameters in each model above. The Akaike Information Criteria \cite[AIC; ][]{1974ITAC...19..716A,2002msmi.BA} provides for a statistical model selection from the above parametric spectral models. Assuming that the model parameter errors are Gaussian distributed, the AIC is 
\begin{equation}
AIC = \chi^2+2 k, 
\end{equation}
where $k$ is the number of free parameters in the model. The particular model with the least AIC is deemed the most likely to best fit the data. If we take the model with the least AIC as the null, the relative importance of model can be gauged by evaluating the difference $\Delta_i = {\rm AIC}_i - {\rm AIC}_{\rm null}$ ($i$ runs through the models excluding the null). With $\Delta_i \leq 2$, both models are indistinguishable, with $4 \leq \Delta_i \leq 7$ the model is considerably less supported, and with $\Delta_i > 10$ the null is still the best fit \cite[e.g.][]{2004smr.BA}.

\section{Results and discussion}


All the X-ray spectra were fit with the two models discussed above: a power law (PL) model which includes an excess absorption column density as a free parameter to identify spectral flattening due to absorption in excess of the Galactic contribution, and, a power law model multiplied by an quadratic form of an exponential roll-off (EXP) to identify spectral flattening due to the intrinsic electron distributions and radiative processes. The model fitting results for all X-ray spectra using {\it XMM--Newton} and {\it Swift} observations are summarized in Table \ref{paratabxmm_2} and \ref{paratabswift} respectively. The results for individual sources and in the context of earlier observations are discussed in Appendix \ref{sampledes}. Few examples of the spectra fitted with models along with their respective ratios are presented in Fig \ref{fig1}. Inferences are drawn by considering results from the fitting of both {\it XMM-Newton} and {\it Swift} X-ray spectra. 


In a majority of the sources (ten of sixteen), we are unable to make a statistically sound or physically motivated distinction between the fit models. An intrinsically shaped particle energy distribution (and hence that of the produced photons) accelerated by the jet is found to shape the spectra (EXP model) in four of the FSRQs: RBS 315, PKS 1830$-$211, QSO B0014$+$810 and PKS 2149$-$306. For these, the spectral index $\Gamma$ is in the range 1.22 to 1.97, and the folding energy of the exponential roll-off $E_f$ is in the range 0.356 keV to 1.127 keV (observer frame). 
Through the identification of a spectral break, the soft X-ray spectral flattening can importantly constrain the lower end of the electron energy distribution to provide inputs on the acceleration mechanisms of relativistic electrons and probe the external Compton (EC) scattering of ambient photons \cite[e.g.][]{2007ApJ...665..980T}. The break is expected at \citep[e.g.][]{2007ApJ...665..980T,2009AdSpR..43.1036F}
\begin{equation}
\nu_{\rm br} = \frac{\nu_{\rm ext} \Gamma^{2}_{b} \gamma^2_{\rm min}}{(1+z)}
\end{equation}
where  $\nu_{\rm ext}$ is the peak frequency of the external radiation field, $\Gamma_{b}$ is the bulk Lorentz factor of the jet and $\gamma_{min}$ is the minimum Lorentz factor of the electron distribution. We assume fiducial values of the above parameters: $\Gamma_{b}$ $\sim$ 10 - 20 \cite[e.g.][]{2019ApJ...874...43L}, $\gamma_{\rm min}$ $\sim$ 1, $\nu_{\rm ext} = 0.41 - 3.75 \times 10^{15}$ Hz \cite[assuming that the accretion disk based emission is the major contributor,][]{2018MNRAS.473.3638G}. With these, the inferred break energy is in the range 0.04 - 1.42 (QSO B0014$+$810), 0.05 - 1.68 (RBS 315), 0.05 - 1.77 (PKS 1830$-$211) and 0.05 - 1.85 (PKS 2149$-$306). These are consistent with the current estimates for these sources (see Tables \ref{paratabxmm_2} and \ref{paratabswift}) and imply that the observed flattening is more likely from jet based intrinsic radiative processes.  

In the remaining two FSRQs, PBC J1656.2$-$3303 and PKS 2126$-$158, the results are mixed. For PBC J1656.2$-$3303, the {\it XMM-Newton} spectrum is better fit with the EXP model while the {\it Swift} spectrum is better fit with the PL model, while the situation is reversed for PKS 2126$-$158, i.e. the {\it XMM} spectrum is better fit with the PL model while the {\it Swift} spectrum is better fit with the EXP model. The analysis is thus indicative of the difficulty in identifying sources particularly where the excess absorption scenario shapes their X-ray spectra.

We now discuss the potential to probe the contributors to the intervening column density for sources with X-ray spectra that may be supportive of the excess absorption scenario.  
The relative contributions to the hydrogen number density from the ICM (composing the galaxy cluster that the blazar is a part of) and the IGM can be disentangled. Assuming that the hydrogen in the IGM along the line of sight has a mean number density \cite[e.g.][and references therein]{2011ApJ...734...26B}
\begin{equation}
n_{0} = 0.67~\Omega_b \frac{3 H^2_0}{8 \pi G m_p} \sim (1.7 \times 10^{-7}~{\rm cm^{-3}}),
\end{equation}
assuming a baryon fraction $\Omega_b = 0.049$ and Hubble constant $H_0 = 67.4$ km s$^{-1}$ Mpc$^{-1}$ \citep{2020A&A...641A...6P}, and for the gravitational constant $G$ and proton mass $m_p$. The integrated line of sight hydrogen column density in the IGM is then \citep{2013MNRAS.431.3159S}
\begin{equation}
N_{H,z} = \frac{n_0 c}{H_0} \int^{z}_{0} \frac{(1+z)^2 dz}{[\Omega_m (1+z)^3+\Omega_\Lambda]^{1/2}},
\end{equation}
where $n_0 c/H_0 = 2.34 \times 10^{21}$ cm$^{-2}$ for the assumed $H_0$; we also take the matter density parameter $\Omega_m = 0.315$ and the vacuum density parameter $\Omega_\Lambda = 1-\Omega_m$ \citep{2020A&A...641A...6P} for the subsequent calculations. 

The ICM number density profile is presumed to be well fit with the $\beta$-model \cite[e.g.][]{1995MNRAS.275..720N}
\begin{equation}
n_{H,c} = n_c \left(1+\left(\frac{r}{r_c}\right)^2\right)^{-3 \beta/2},
\end{equation}
where $n_c$ is the characteristic number density of the galaxy cluster with an associated size of $r_c$, $r$ is the distance to the center and $\beta > 0$ is an index shaping the number density distribution. The integrated hydrogen column density \cite[using the simplification $\left<\beta\right> \approx 0.5$;][]{2016MNRAS.459..366G} in the ICM is then
\begin{align}
N_{H,c} &= \int^{r_c}_0 n_{H,c} dr \\ \nonumber
&= (0.26 \times 10^{22}~{\rm cm^{-2}})\left(\frac{n_c}{10^{-3}{\rm cm^{-3}}}\right) \left(\frac{r_c}{\rm Mpc}\right).
\end{align}

For blazars potentially hosting intervening X-ray absorption components, their excess column density can be compared based on $N_{H} = N_{H,z}+N_{H,c}$. As an illustrative example, we assume that the FSRQs PBC J1656.2$-$3303 and PKS 2126$-$158 tentatively support this scenario. 
The ICM size $r_c$ is fixed at 1 Mpc, a fiducial size for galaxy clusters. The formulation is used to estimate the ICM number density $n_c$ for the blazars where $N_H > N_{H,z}$, 
and is presented in Table \ref{NHICM}. 
The inferred number density in the case of PBC J1656.2$-$3303 is similar to the expectation of 10$^{-3}$ -- 10$^{-2}$ cm$^{-3}$ \cite[e.g.][]{1986RvMP...58....1S,2013PhR...533...69C}. In the case of PKS 2126$-$158, we are unable to make this estimate as $N_H < N_{H,z}$ for this source.

Owing to the extremely limited sample size, the difficulty in the direct observational identification of galaxy clusters at high redshifts and limitations from the data quality affecting the estimated excess column density (spectral fitting), this may only be a plausible conclusion. This then requires continued monitoring with higher resolution X-ray spectroscopy, possibly supplemented by multi-wavelength spectroscopic observations that can better quantify the relative contributions to the line of sight column density. Additionally, owing to large parameter uncertainties and a limited sample size, it is difficult to provide further clarification on the nature of the EC contributors and the electron indices before and after the break in the FSRQs that indicate a low energy spectral flattening attributable to intrinsic jet based processes.

\begin{table*}
\begin{center}
\caption{Estimates of number density in a sub-sample of blazars indicating a statistically or physically identified excess column density based on the X-ray spectral fits.}
\begin{tabular}{cccccc} \hline \hline
Blazar & Redshift & $\left<N_H\right>$ & $N_{H,z}$ & $N_{H,c}$ & $n_c$ \\ 
       & $z$      & (excess) & (IGM) & (ICM) & (ICM)\\
       & & ($\times 10^{22}$ cm$^{-2}$) & ($\times 10^{22}$ cm$^{-2}$) & ($\times 10^{22}$ cm$^{-2}$) & ($\times 10^{-3}$ cm$^{-3}$) \\ \hline
PBC J1656.2$-$3303 & 2.40 & 11.4 & 1.3 & 10.1 & 39.0 \\
PKS 2126$-$158 & 3.27 & 1.4 & 2.0 & - & -\\ \hline      
\end{tabular}   
\label{NHICM}
\end{center}
\end{table*}

\section*{Acknowledgements}
We thank the referee for her/his comments and suggestions that have helped clarify the analysis and the consequent novel aspects of our work. We thank the editor for a careful reading and for helpful suggestions on the presentation. H.G. thanks Dr. J.C. Pandey for his help and discussion in this work. This research is based on observations obtained with XMM$-$Newton, an ESA science mission with instruments and contributions directly funded by ESA member states and NASA. HG acknowledges the financial support from the Department of Science \& Technology, India through INSPIRE faculty award IFA17-PH197at ARIES, Nainital. 

\appendix

\section{Notes on individual blazars} 
\label{sampledes}

We discuss the previous X-ray observations and inferences drawn in literature. Our combined assessment of the {\it XMM-Newton} and {\it Swift} is then presented and compared with the reports from literature.\\

{\bf QSO B0014$+$810:} \cite{2005MNRAS.364..195P} study the {\it XMM-Newton} observation of 2001 August 23 and infer a photon index $\Gamma$ of 1.61, an X-ray luminosity of $5.94 \times 10^{47}$ erg s$^{-1}$ and an excess absorption column density $N_H$ of $1.82\times 10^{22}$ cm$^{-2}$. \cite{2009MNRAS.399L..24G} re-analyze the {\it XMM-Newton} observation during 2001 and infer a harder photon index $\Gamma$ of 1.44 and a $N_H$ of $6.00 \times 10^{21}$ cm$^{-2}$, attributable to the screening of flaring backgrounds and improved calibration; the study finds a better fit with the broken power law model (including Galactic absorption) which provides an index $\Gamma_1$ of 1.1 below the break energy $E_b$ at 1.0 keV and $\Gamma_2$ of 1.43 above $E_b$. This suggests a spectrum shaped by intrinsic radiative processes. \cite{2018A&A...616A.170A} study the {\it XMM-Newton} observation of 2001 August 23 and find a $\Gamma$ in the range of 1.49 - 1.77 and excess absorption $N_H$ of $< 0.54 \times 10^{22}$ cm$^{-2}$ with an inability to distinguish between excess absorption and intrinsic radiative processes; they also find a warm IGM with a temperature of $> 6.9 \times 10^6$ K and number density of $\approx 10^{-7}$ cm$^{-3}$ in the context of the IGM contribution to the excess absorption along the line of sight. 

Our analysis of the X-ray spectra suggests a preference for the EXP model with a $\Gamma$ of 2.0 and $E_{f}$ of 0.7 keV.\\

{\bf RBS 315:} A power law with excess absorption is found to better fit the spectrum (over a broken power law model) and yields a photon index $\Gamma$ of 1.23 and $N_H$ of 1.6 - 2.2$\times 10^{22}$ cm$^{-2}$ indicating a strong absorption component \citep{2005A&A...442L..53P}. \cite{2005MNRAS.364..195P} infer a similar $\Gamma$ of 1.22 and $N_H$ of $1.74 \times 10^{22}$ cm$^{-2}$ from the same {\it XMM-Newton} observation. \cite{2007ApJ...665..980T} study the {\it Suzaku} X-ray spectrum (0.3 - 50 keV) from 2006 July 25 - 27 and infer that the spectrum is better described by intrinsic radiative processes in the jet. 
\cite{2016MNRAS.462.1542S} study the {\it Swift} and {\it NuSTAR} based observations, both from 2014 December 24 and 2015 January 18 in the context of modelling the broadband spectral energy distribution and find a broken power law with indices $\Gamma_1$ of $1.06 - 1.10$ and $\Gamma_2$ of $1.56 - 1.69$, and $E_b$ of $4.69 - 4.77$ keV and accounting for Galactic absorption alone to be a better description of the combined 0.5 - 50 keV X-ray spectrum. \cite{2018A&A...616A.170A} study the {\it XMM-Newton} based X-ray spectra from three epochs (2003 July 25, 2013 January 13 and 15) and find that the log-parabola with excess absorption model provides the best description and report an $N_H$ of $0.75 \times 10^{22}$ cm$^{-2}$. The recent study of \cite{2019ApJ...882..130B} includes spectra from {\it XMM-Newton}, {\it Swift} and {\it NuSTAR} observations (0.3 - 70 keV) spanning between 2013 January - 2015 January; the fitted column densities $N_H$ are consistently in the range $3.2 - 4.4 \times 10^{22}$ cm$^{-2}$ in all epochs with $\Gamma \approx 1.4 - 1.6$ and indicate a possible absorber in the host galaxy. 

Our analysis of three {\it XMM-Newton} spectra suggests a preference for the EXP model with $\Gamma$ of 1.26 -- 1.48 and $E_{f}$ of 0.36 -- 0.40 keV which is attributed to jet based intrinsic physical processes, consistent with the study of \cite{2007ApJ...665..980T}.\\

{\bf PMN 0525$-$3343:} The analysis of the {\it ASCA} and {\it BeppoSAX} X-ray spectra (restricted to the 0.8 - 10 keV energy range) from 1999 and 2000 respectively consistently show a spectral flattening which is best understood in terms of a power law with excess absorption attributable to the host galaxy, with an index $\Gamma$ of $\approx$ 1.6 - 1.7 and $N_H$ of $\approx 10^{23}$ cm$^{-2}$ \citep{2001MNRAS.323..373F}. \cite{2004MNRAS.350..207W} study the {\it XMM-Newton} observations from 2001 (two epochs) and 2003 (six epochs) and confirm the spectral flattening and attribute it to a warm ionized absorber in the host galaxy with a column density $N_H$ of $2 - 3 \times 10^{22}$ cm$^{-2}$; their power law model (including the excess absorption) yields an index $\Gamma$ of 1.64. \cite{2005MNRAS.364..195P} study the {\it XMM-Newton} observation of 2001 September 15 and infer a photon index $\Gamma$ of 1.73, a luminosity of $3.51 \times 10^{47}$ erg s$^{-1}$ and an excess absorption column density $N_H$ of $1.91\times 10^{22}$ cm$^{-2}$, the latter constraint likely due to better data quality with {\it XMM-Newton} to explain the deficit compared to the estimate of \citep{2001MNRAS.323..373F}. \cite{2005A&A...433.1163D} study the 2001 {\it BeppoSAX} spectrum and find that a simple power law with index $\Gamma$ of 1.58 and including only the Galactic absorption provides a reasonable fit. \cite{2018A&A...616A.170A} study the {\it XMM-Newton} based X-ray spectra from three epochs (2001 September 09, 2003 February 14 and 24) and find a degeneracy between a broken power law model (including Galactic absorption) with indices $\Gamma_1$ of $0.64 - 1.33$ and $\Gamma_2$ of 1.6, and a power law including excess absorption model with index $\Gamma$ of 1.6 and $N_H$ of $0.93 \times 10^{22}$ cm$^{-2}$.  

Our analysis of the X-ray spectra does not favor any particular model and is consistent with the above mixed results, necessitating new higher resolution X-ray spectroscopic observations.\\

{\bf PKS 0524$-$460:} The BAT AGN spectroscopic survey \citep{2017ApJS..233...17R} employs the X-ray spectrum in the 0.3 — 150 keV energy range composed of observations from {\it XMM-Newton}, {\it Swift}, {\it ASCA}, {\it Chandra} and {\it Suzaku}; a power law with excess absorption fit to the extended spectrum provides an index $\Gamma$ of 1.28 and indicates an unobscured source. 

Our analysis of the X-ray spectra does not favor any particular model, necessitating new higher resolution X-ray spectroscopic observations.\\

{\bf PKS 0528$+$134:} \cite{1999A&A...348...63G} study the {\it BeppoSAX} observations (0.1 - 10 keV; 15 - 200 keV) made in eight epochs during 1997 February 21 - March 11; a power law with excess absorption fit provides a $\Gamma$ of $0.37 - 0.48$ and $N_H$ of $3.9 - 5 \times 10^{21}$ cm$^{-2}$, consistent with results from the ASCA observations (0.5 - 10 keV) of the source during 1995 March \citep{1997ApJ...474..639S}. \cite{2005A&A...433.1163D} study the BeppoSAX observations (covering 0.1 - 300 keV energy range) made in eight epochs during 1997 February and March, and find that a power law (including only the Galactic absorption) is best fit with an index $\Gamma$ in the range $1.12 - 1.62$. \cite{2011ApJ...735...60P} study four {\it XMM-Newton} observations from 2009 September 8 - 14 in the context of a multi-wavelength campaign on the source during a quiescent state; a power law with excess absorption fit provides an index $\Gamma$ in the range $1.58 - 1.61$ and $N_H$ in the range $0.12 - 0.17 \times 10^{22}$ cm$^{-2}$. The BAT AGN spectroscopic survey \citep{2017ApJS..233...17R} X-ray spectrum (0.3 - 150 keV) is fit with a power law model with index $\Gamma$ of 1.60 and absorbing column density $N_H$ of $3.7 \times 10^{22}$ cm$^{-2}$. \cite{2018A&A...616A.170A} study the {\it XMM-Newton} based X-ray spectra from three epochs (2009 September 10, 11 and 14) and find that a power law with excess absorption provides the best fit; this yields an index $\Gamma$ of $\approx 1.6$ and $N_H$ of $1.45 \times 10^{22}$ cm$^{-2}$. The results indicate that an excess absorption is likely, though with uncertainties in the column density.


Our analysis of the X-ray spectra does not favor any particular model and is contradictory to the above inferences from literature. This likely owes to the differing spectral models employed and necessitates new higher resolution X-ray spectroscopic observations.\\

{\bf PKS 0537$-$286:} \cite{1997ApJ...478..492C} study the X-ray observations from {\it ROSAT} (0.1 - 2.4 keV; 1992 September 28) and {\it ASCA} (0.5 - 10 keV; 1994 March 12) and find that the combined X-ray spectrum fit with a power law with excess absorption model yields a photon index $\Gamma$ of 1.4 and excess column density $N_H$ of $1.7 \times 10^{21}$ cm$^{-2}$ which is found to marginally exceed the Galactic contribution. \cite{2001A&A...365L.116R} study the {\it XMM-Newton} observations from 1999, finding the source to be highly X-ray luminous ($2 \times 10^{47}$ erg s$^{-1}$); they find that a power law fit (including Galactic absorption) with an index $\Gamma$ of $\approx 1.3$ best describes the spectrum, without the additional necessity for an absorber at the host galaxy redshift, and note that the X-ray emission likely originates from inverse-Compton processes. \cite{2010A&A...509A..69B} study the source in the context of a multi-wavelength campaign during 2006 - 2008, including the 1999 and 2005 {\it XMM-Newton} observations; a power law with excess absorption fit to the {\it XMM-Newton}, {\it RXTE} and {\it Swift} (a combined 0.6 - 150 keV) based X-ray spectra provide a $\Gamma$ ranging between $1.3 - 1.7$ and $N_H$ of $\approx 5 \times 10^{21}$ cm$^{-2}$, consistent with the existence of a time varying (years timescale) local absorber in the host galaxy. \cite{2018A&A...616A.170A} study the {\it XMM-Newton} based X-ray spectra from two epochs (2000 March 19, 2005 March 20) and find that the broken power law model provides a better fit (though the log-parabola model with excess absorption may not be entirely ruled out), with indices $\Gamma_1$ of 1.35 and $\Gamma_2$ of 1.18, and a break energy $E_b$ of 2.39 keV. The mixed results suggest that the spectral flattening may be shaped by either a varying absorption column (though weak) or intrinsic radiative processes. 

Our analysis of the X-ray spectra does not favor any particular model, necessitating new higher resolution X-ray spectroscopic observations.\\

{\bf 1ES 0836$+$710:} \cite{1997ApJ...478..492C} study the X-ray observations from {\it ROSAT} (0.1 - 2.4 keV; 1992 March 23, November 02) and {\it ASCA} (0.5 - 10 keV; 1995 March 17) and find that the X-ray spectra indicate a photon index $\Gamma$ of $\approx 1.5$ and $N_H$ in the range $0.33 - 1.17 \times 10^{22}$ cm$^{-2}$, moderately larger than the Galactic contribution hinting at the presence of an absorber in the host galaxy. \cite{2000ApJ...543..535T} study the BeppoSAX (0.1 - 200 keV) X-ray spectrum obtained on 1998 May 27 - 28 which is fit well by a power law with excess absorption in the host galaxy and yields a $\Gamma$ of 1.33 and $N_H$ of $6.6 \times 10^{21}$ cm$^{-2}$, though it is noted that a broken power law (including a fixed Galactic absorption) with a relatively flat lower energy slope provides a comparable fit, in agreement with the study of \cite{1997ApJ...478..492C}. \cite{2000MNRAS.316..234R} study the {\it ASCA} observations from 1995, finding that a power law with excess absorption fit provides an index $\Gamma$ of 1.41, $N_H$ of $9 \times 10^{21}$ cm$^{-2}$ and an integrated 2 - 10 keV luminosity of $2.2 \times 10^{47}$ erg s$^{-1}$. \cite{2005MNRAS.364..195P} study the {\it XMM-Newton} observations from 2001 April 13 and find that a broken power law model (including only the Galactic absorption) provides a better fit to the X-ray spectrum and provides indices $\Gamma_1$ of 0.59, $\Gamma_2$ of 1.36, and an energy break $E_b$ at 1.62 keV. \cite{2006A&A...453..829F} study the {\it XMM-Newton} observations from 2001 April 12, finding that a power law with excess absorption is the best fit model and provides an index $\Gamma$ of 1.38 and $N_H$ of $1.4 \times 10^{21}$ cm$^{-2}$. The BAT AGN spectroscopic survey \citep{2017ApJS..233...17R} X-ray spectrum (0.3 - 150 keV) is fit with a power law model with index $\Gamma$ of 1.42 and absorbing column density $N_H$ of $2.0 \times 10^{21}$ cm$^{-2}$.

Our analysis of the X-ray spectra does not favor any particular model, necessitating new higher resolution X-ray spectroscopic observations.\\

{\bf RX J1028.6$-$0844:} The source received attention owing to being one of the earliest with a reported spectral flattening at low X-ray energies, attributed to possibly the densest column density in an absorber intrinsic to the host galaxy \cite{2000ApJ...545..625Y}; the 1999 November 25 {\it ASCA} observations (0.5 - 10 keV) based X-ray spectrum is best fit by a power law including excess absorption, yielding a photon index $\Gamma$ of 1.72 and $N_H$ ranging between $2.1 \times 10^{23} - 1.6 \times 10^{24}$ cm$^{-2}$ depending on the assumed metallicity of the absorbing gas. The study of \cite{2004AJ....127....1G} however analyzes the higher resolution {\it XMM-Newton} based X-ray spectrum of 2002 May 15 to infer that the reported {\it ASCA} based $N_H$ may be vastly overestimated, and is consistent with the Galactic contribution of $4.6 \times 10^{20}$ cm$^{-2}$ indicating the absence of a strongly absorbing column of gas in the host galaxy. \cite{2005MNRAS.364..195P} study the {\it XMM-Newton} observation from 2003 June 13, finding that a broken power law (including only the Galactic absorption) provides a marginally better fit to the X-ray spectrum, with indices $\Gamma_1$ of 1.12, $\Gamma_2$ of 1.47, and a break energy $E_b$ at 6.06 keV. 

Our analysis of the X-ray spectra does not favor any particular model, necessitating new higher resolution X-ray spectroscopic observations.\\

{\bf PKS 1127$-$145:} The FSRQ is an extremely radio bright GHz peak spectrum (GPS) source (based on the synchrotron peak and turnover in the GHz radio frequencies) with a 5 GHz flux density of 3.82 Jy \citep{1998A&AS..131..303S}. It has a one sided X-ray jet emission at $\approx$ 300 kiloparsec away from the central engine \citep{2002ApJ...570..543S,2007ApJ...657..145S}. 
GPS sources are morphologically compact ($\leq$ 1 kiloparsec) with their jet either developing (young AGN scenario) or frustrated owing to a dense surrounding medium \cite[e.g.][]{1998PASP..110..493O}. Clues from X-ray observations can help distinguish between these scenarios.

\cite{2005A&A...433.1163D} study the {\it BeppoSAX} observation from 1999 June 12, finding that a power law with $N_H$ fixed to the Galactic contribution best fits the X-ray spectrum and yields a photon index $\Gamma$ of 1.42. \cite{2006A&A...453..829F} study the 2002 July 01 {\it XMM-Newton} based observation and find the X-ray spectrum to be well fit by a broken power law with excess absorption model, yielding indices $\Gamma_1$ of 1.40, $\Gamma_2$ of 1.22, a break energy $E_b$ at 2.7 keV and an excess column density $N_H$ of $1.2 \times 10^{21}$ cm$^{-2}$ associated with the host galaxy. 

Our analysis of the X-ray spectra does not favor any particular model, necessitating new higher resolution X-ray spectroscopic observations. These can especially help disentangle the relative contributions from the local environment and that intrinsic to the X-ray jet.\\  


{\bf PKS 1406$-$076:} \cite{2006A&A...453..829F} study the 2003 July 05 and August 08 {\it XMM-Newton} based observation and find the X-ray spectra to be well fit by a power law with index $\Gamma$ of 1.59 and consistent with the Galactic absorption alone. \cite{2013ApJS..207...28S} study the {\it Swift-XRT} (in the 0.3 - 10 keV energy range) monitoring observations between 2004 - 2012 in the context of a study of {\it Fermi} $\gamma$-ray sources, and find that the X-ray spectrum fit with a power law including excess absorption yields an average photon index $\Gamma$ of 3.0 and $N_H$ of $3 \times 10^{21}$ cm$^{-2}$.

Our analysis of the X-ray spectra does not favor any particular model, necessitating new higher resolution X-ray spectroscopic observations.\\

{\bf 7C 1428$+$4218:} \cite{2000MNRAS.315L..23B} study the {\it ROSAT} observations from 1998 December 11 and 17, finding evidence that the soft X-ray absorption originates from an excess column density $N_H \approx 1.5 \times 10^{22}$ cm$^{-2}$ in the host galaxy. 
\cite{2001MNRAS.324..628F} study the {\it BeppoSAX} observations (in the 0.4 - 10 keV energy range) from 1999 February 4 - 7, finding that fitting the X-ray spectrum with a power law including excess absorption yields a photon index $\Gamma$ of $\approx 1.5$ and a highly ionized gas with column density $N_H$ of $\approx 10^{23}$ cm$^{-2}$. \cite{2005MNRAS.364..195P} study the {\it XMM-Newton} observations from 2003 January 17, finding that the power law with excess absorption and broken power law (including only the Galactic absorption) models provide comparable fits; the former yields a photon index $\Gamma$ of 1.75 and $N_H$ of $1.62 \times 10^{22}$ cm$^{-2}$ while the latter yields indices $\Gamma_1$ of 0.85 and $\Gamma_2$ of 1.71 and a break energy $E_b$ at 3.71 keV (broken power law). The BAT AGN spectroscopic survey \citep{2017ApJS..233...17R} X-ray spectrum (0.3 - 150 keV) is fit with a power law model with index $\Gamma$ of 2.00 and absorbing column density $N_H$ of $2.3 \times 10^{20}$ cm$^{-2}$, only marginally larger than the Galactic contribution. \cite{2018A&A...616A.170A} study the {\it XMM-Newton} based X-ray spectra from two epochs (2003 January 17, 2005 June 05) and find that the broken power law (including only the Galactic absorption), and a power law and log-parabola with excess absorption models provide comparable fits; the broken power law yields indices $\Gamma_1$ of $\approx 0.7$, $\Gamma_2$ of $\approx 1.5 - 1.7$, and break energy $E_b$ at 0.59 keV, while power law and log-parabola models yield an index $\Gamma$ of $1.5 - 1.7$ and $N_H$ of $1.9 - 2.3 \times 10^{22}$ cm$^{-2}$.   

Our analysis of the X-ray spectra does not favor any particular model, necessitating new higher resolution X-ray spectroscopic observations.\\

{\bf GB 1508$+$5714:} \cite{1995AJ....110.1551M} study the {\it Einstein} observatory (0.4 - 4 keV) based archival observations, finding that an excess $N_H$ of $> 10^{22}$ cm$^{-2}$ is required to fit the X-ray spectrum. \cite{1997ApJ...484L..95M} study the {\it ASCA} (0.5 - 10 keV) observations, finding the source to be highly luminous ($\approx 10^{47}$ erg s$^{-1}$) finding however that the X-ray spectrum is adequately fit by a power law with index $\Gamma$ of $\approx$ 1.4 and including only the Galactic absorption. \cite{2003MNRAS.346L...7Y} report a photon index $\Gamma$ of 1.92 based on a power law (with a fixed Galactic absorption) fit to the 0.5 - 8 keV X-ray spectrum obtained by {\it Chandra}. \cite{2005MNRAS.364..195P} study the {\it XMM-Newton} observations from 2002 May 11, finding the X-ray spectrum to be best fit by a broken power law (including only the Galactic absorption) with indices $\Gamma_1$ of 1.62, $\Gamma_2$ of 1.08, and a break energy $E_b$ at 17.12 keV. \cite{2006MNRAS.368..985Y} study the same {\it XMM-Newton} based X-ray spectrum from 2002 May 11, finding evidence for absorption in the host galaxy based on a column density $N_H$ of $< 3 \times 10^{21}$ cm$^{-2}$ and a power law index $\Gamma$ of $\approx 1.5$.   

Our analysis of the X-ray spectra does not favor any particular model, necessitating new higher resolution X-ray spectroscopic observations.\\

{\bf PBC J1656.2$-$3303} The source was identified in a Swift/BAT hard X-ray (14 - 200 keV) survey during 2004 December - 2005 September, with a reported preliminary photon index $\Gamma$ of 1.26 \citep{2006ATel..799....1O}. The Swift/BAT based X-ray spectrum from 2006 June 09 is found to be fit well by a power law with excess absorption, yielding a photon index $\Gamma$ of 1.38 and an excess column density $N_H$ of $2.21 \times 10^{21}$ cm$^{-2}$ \citep{2006ATel..835....1T}. \cite{2008A&A...480..715M} study the optical, X-ray and radio properties confirming its blazar nature; 
Swift/XRT observations on 2006 June 9 and 13 and Integral/IBIS observations between 2002 October and 2006 April result in a 0.7 - 200 keV continuum which is fit well by both a broken power law (including only Galactic absorption) with $\Gamma_1$ of 0.86, $\Gamma_2$ of 1.81, and $E_b$ at 2.7 keV, and a power law including excess absorption with $\Gamma$ of 1.64, and $N_H$ of $6.7 \times 10^{22}$ cm$^{-2}$. \cite{2018A&A...616A.170A} study the {\it XMM-Newton} observation from 2009 September 11, and NuSTAR and Swift/XRT based X-ray spectra from 2015 September 27, finding that the broken power law model (including Galactic absorption) provides a better fit, and yields indices $\Gamma_1$ of 0.88, $\Gamma_2$ of 1.19, and a break energy $E_b$ at $1.44 - 2.24$ keV; they also find a warm IGM with a temperature of $5.4 \times 10^5$ K and number density of $\approx 10^{-7}$ cm$^{-3}$ in the context of the IGM contribution to the excess absorption along the line of sight. 


Our analysis of the {\it XMM-Newton} spectrum favors the EXP model with a $\Gamma$ of 1.29 and a folding energy of 0.89 keV based. The analysis of the {\it Swift} spectrum however favors the PL model with a $\Gamma$ of 1.6 and an excess column density $N_H$ of 15 $\times$ 10$^{22}$ cm$^{-2}$. The mixed results necessitate new higher resolution X-ray spectroscopic observations.\\

{\bf PKS 1830$-$211:} \cite{1997ApJ...484..140M} study the {\it ROSAT} (0.1 - 2.4 keV) observations during 1993 September 15, finding evidence for an absorbing column of gas exceeding the Galactic contribution; a power law fit gives an index $\Gamma$ of $\approx 1.3$ and a $N_H$ of $3.5 \times 10^{22}$ cm$^{-2}$, the latter being attributable to the intervening lensing galaxy at $z = 0.886$. \cite{2001ApJ...551..929O} study eight {\it ASCA} observations (0.5 - 10 keV) from 1999 September 11 to October 15, finding a spectral variability which may result from a varying absorption or in terms of micro-lensing of two spectral components with differing absorption and varying brightness; a two component power law model (including a fixed Galactic contribution) fit yields a photon index $\Gamma$ of $\approx 1.3 - 1.5$ and $N_H$ of $7.5 \times 10^{22}$ cm$^{-2}$ and $< 1.5 \times 10^{22}$ cm$^{-2}$ at the redshift of the intervening lensing galaxy. The study of \cite{2005A&A...438..121D} confirms that the observed spectral flattening is attributable to an excess absorption, finding that the X-ray spectra based on {\it Chandra} (0.3 - 5 keV only) and {\it INTEGRAL} (20 - 100 keV) observations from 2000 June 26 - 27 and 2001 June 25, and 2003 February 28 to October 10 respectively yields a power law index $\Gamma$ of $\approx 1.1$; it is noted that in addition to an $N_H$ of $\approx 10^{22}$ cm$^{-2}$ in the lensing galaxy, there could be a contribution of $\approx 10^{23}$ cm$^{-2}$ from the blazar host galaxy. \cite{2008AJ....135..333D} study two observations with {\it Chandra} (0.35 - 8 keV) from 2000 June 26 and 2001 June 25, and three observations with {\it XMM-Newton} (0.35 - 10 keV) from 2004 March 10, 24 and May 05, finding that the spectral variability is attributable to a varying absorption; fits to the X-ray spectrum yield a photon index $\Gamma$ of $\approx 1.0 - 1.3$, and enable the identification of the absorption column density contributions from the Galaxy ($N_H$ of $2.2 \times 10^{21}$ cm$^{-2}$), the lensing galaxy ($N_H$ of $\approx 2 \times 10^{22}$ cm$^{-2}$) and that intrinsic to the blazar host galaxy ($N_H$ of upto $\approx 30 \times 10^{22}$ cm$^{-2}$); the large variations in column density are attributed to polar outflows from the central engine. These clues raise the possible connection between the fast outflows and the driving central engine as active components in feedback with the host galaxy and the large scale environment.

Our analysis of the {\it XMM-Newton} spectra suggests a tentative preference for the EXP model with a $\Gamma$ of 1.22 and a folding energy of 1.13 keV which is attributed to jet based intrinsic physical processes.\\

{\bf PKS 2126$-$158:} This FSRQ is also a GPS source and is of similar interest as PKS 1127$-$145 in the context of the synchrotron turnover originating from intrinsic jet based processes (SSA) or from the surrounding external environment (FFA), connected to the physical size of the GPS radio lobes and their dynamic evolution \cite[e.g.][]{2012ApJ...760...77A}.

Early {\it ROSAT} (0.1 - 2.4 keV) observations from 1991 May 9 and 1992 November 12 - 13 show evidence for the presence of an absorber in the host galaxy owing to a column density $N_H$ of $1.45 \times 10^{22}$ cm$^{-2}$ which exceeds the Galactic contribution, and provides a power law index $\Gamma$ of $\approx 0.5$ \citep{1994ApJ...422...60E}. \cite{1994PASJ...46L..43S} study the {\it ASCA} (0.5 - 10 keV) observations from 1993 May 16 - 17, finding that the X-ray spectrum is well fit either by a power law with an intervening absorber at low redshift (including Galactic absorption) or a broken power law (with fixed Galactic absorption) model with the latter being moderately better; the former provides an index $\Gamma$ of 1.62, and $N_H$ of $\approx 10^{21}$ cm$^{-2}$ at a redshift $z = 0.027$ while the latter provides indices $\Gamma_1$ of 1.11, $\Gamma_2$ of 1.70, and a break energy $E_b$ at 1.91 keV. \cite{1997ApJ...478..492C} study the {\it ROSAT} observations on four epochs from 1991 May 08 - 1993 May 13 and {\it ASCA} observations on 1993 May 16 and modelling the combined X-ray spectra provide better fits but result in a degeneracy between them; a power law model with fixed Galactic absorption and free excess absorption (at the host galaxy redshift) provides $\Gamma$ of $\approx 1.6$ and $N_H$ of $1.16 \times 10^{22}$ cm$^{-2}$, and a broken power law model (with a fixed Galactic absorption) provides a $\Gamma_1$ of 1.02, $\Gamma_2$ of 1.62, and a break energy $E_b$ at 1.26 keV. \cite{2003A&A...409...57F} study the {\it BeppoSAX} (0.1 - 100 keV) observations from 1999 May 24 - 28, finding that the X-ray spectrum is better fit by a broken power law (including a fixed Galactic absorption and free excess absorption at the host galaxy redshift) which yields indices $\Gamma_1$ of $0.49 - 0.73$, $\Gamma_2$ of $1.5 - 3.5$, a break energy $E_b$ at $5 - 36$ keV, and an excess absorption column density $N_H$ of $0.13 \times 10^{22}$ cm$^{-2}$. \cite{2003A&A...402..465F} study the {\it XMM-Newton} observation from 2001 May 01, finding that the PN detector based X-ray spectrum fit with a power law including excess absorption yields an index $\Gamma$ of 1.47 and $N_H$ of $1.40 \times 10^{22}$ cm$^{-2}$. \cite{2005MNRAS.364..195P} study the same {\it XMM-Newton} observation from 2001 May 01, finding that a power law with excess absorption model provides the best fit and yields an index $\Gamma$ of 1.44 and $N_H$ of $1.19 \times 10^{22}$ cm$^{-2}$. \cite{2018A&A...616A.170A} study the same {\it XMM-Newton} observation from 2001 May 01, and find that the power law and log parabola models with excess absorption provide better fits to the X-ray spectrum, with an index $\Gamma$ of 1.45, and $N_H$ of $1.24 - 1.38 \times 10^{22}$ cm$^{-2}$.


Our analysis of the {\it XMM-Newton} spectrum favors the PL model with a $\Gamma$ of 1.4 and an excess column density of 1.1 $\times$ 10$^{22}$ cm$^{-2}$. The analysis of the {\it Swift} spectrum however favors the EXP model with a $\Gamma$ of 2.3 and a folding energy of 1.4 keV. The mixed results necessitate new higher resolution X-ray spectroscopic observations.\\

{\bf PKS 2149$-$306:} \cite{1996A&A...307....8S} study the {\it ASCA} (0.5 - 10 keV) observations during 1994 October 26, finding that a power law with excess absorption fit to the X-ray spectrum yields an index $\Gamma$ of $1.54 - 1.57$, and $N_H$ of $< 1.25 \times 10^{21}$ cm$^{-2}$. \cite{1997ApJ...478..492C} and \cite{2000MNRAS.316..234R} study the same {\it ASCA} 1994 October 26 observation, finding that a power law with excess absorption fit to the X-ray spectrum yields an index $\Gamma$ of $1.49 - 1.54$, and a better constrained $N_H$ of $6.3 - 8.3 \times 10^{21}$ cm$^{-2}$, higher than the estimate of \cite{1996A&A...307....8S}. \cite{2005A&A...433.1163D} and \cite{2006ApJ...642..113G} study the {\it BeppoSAX} (0.1 - 100 keV) observations from 1997 October 31, finding that a power law model (including only the Galactic absorption) best fits the X-ray spectrum and provides an index $\Gamma$ of $1.35 - 1.37$, hinting at less evidence for an absorbing column intrinsic to the blazar host galaxy. \cite{2003A&A...402..465F} study the {\it XMM-Newton} observations from 2001 May 01, finding that the PN detector based X-ray spectrum fit with a power law including excess absorption yields an index $\Gamma$ of 1.53 and $N_H$ of $2.94 \times 10^{20}$ cm$^{-2}$, only marginally higher than the Galactic contribution of $2.10 \times 10^{20}$ cm$^{-2}$ indicating that the evidence for an absorbing medium in the host galaxy is not very strong. \cite{2005MNRAS.364..195P} study the same {\it XMM-Newton} observations from 2001 May 01, finding that a power law with excess absorption fit yields an index $\Gamma$ of 1.47 and $N_H$ of $< 3 \times 10^{20}$ cm$^{-2}$, consistent with the study of \cite{2003A&A...402..465F}. \cite{2007ApJ...669..884S} study the {\it Swift} (0.3 - 150 keV) observations from 2005 December 10 and 13, finding that the X-ray spectrum fit by a power law with excess absorption yields an index $\Gamma$ of 1.50 and a $N_H$ of $2.5 \times 10^{21}$ cm$^{-2}$; this $N_H$ is consistent with the {\it XMM-Newton} based inference of it marginally exceeding the Galactic contribution. \cite{2013ApJ...774...29E} study the {\it XMM-Newton} observation from 2001 May 01, finding that both a power law with excess absorption and a broken power law (including a fixed Galactic absorption) provide comparable fits; the former provides a $\Gamma$ of 1.61 and $N_H$ of $8 \times 10^{20}$ cm$^{-2}$ while the latter provides a $\Gamma_1 - \Gamma_2$ of $-0.1$, and a break energy $E_b$ at 3.11 keV. \cite{2018A&A...616A.170A} study the {\it NuSTAR} based X-ray spectra from two epochs (2013 December 17 and 2014 April 18) and find that the broken power law (including only the Galactic absorption), and a power law and log-parabola with excess absorption models provide comparable fits; the broken power law yields indices $\Gamma_1$ of $1.45$, $\Gamma_2$ of $1.16$, and break energy $E_b$ at 6.24 keV, while power law and log-parabola models yield an index $\Gamma$ of $1.35 - 1.46$ and $N_H$ of $< 6 \times 10^{20}$ cm$^{-2}$; they also find a warm IGM with a temperature of $8.7 \times 10^6$ K and number density of $\approx 10^{-7}$ cm$^{-3}$ in the context of the IGM contribution to the excess absorption along the line of sight. The study of \cite{2009A&A...496..423B} infers a spectral roll-off at energies $<$ 1 keV from the {\it Swift} spectra during 2005, and attribute this to intrinsic processes in the jet owing to the onset of inverse Compton emission.


Our analysis of the {\it Swift} spectrum suggests a preference for the EXP model with a $\Gamma$ of  1.59 and a folding energy of 0.51 keV. This is consistent with the inference of \cite{2009A&A...496..423B}, with a consequent attribution of the spectral flattening to intrinsic jet based physical processes.

\end{document}